\documentclass[aps, superscriptaddress, preprintnumbers,reprint]{revtex4-1}

\usepackage{color}
\usepackage{array}
\usepackage{booktabs}
\usepackage[table]{xcolor}
\usepackage{courier}
\usepackage{amsmath,amssymb,amsthm}
\usepackage{mathtools}
\usepackage{cancel}

\usepackage{framed}
\usepackage{graphicx}
\usepackage{subfigure}
\usepackage{appendix}

\usepackage{ulem}

\newcommand{\va}{\boldsymbol{\mathsf{a}}}

\newcommand{\vk}{\boldsymbol{k}}

\ifx \vr\undefined
 \newcommand{\vr}{\boldsymbol{r}} 
\else
 \renewcommand{\vr}{\boldsymbol{r}}
\fi

\newcommand{\vp}{\boldsymbol{p}}
\newcommand{\vq}{\boldsymbol{q}}

\newcommand{\vx}{\boldsymbol{x}}

\newcommand{\vz}{\boldsymbol{z}}

\newcommand{\vF}{\boldsymbol{\mathsf{F}}}
\newcommand{\vFQ}{\boldsymbol{\mathsf{FQ}}}

\newcommand{\vT}{\boldsymbol{\mathsf{T}}}
\newcommand{\vS}{\boldsymbol{\mathsf{S}}}
\newcommand{\vQ}{\boldsymbol{\mathsf{Q}}}

\newcommand{\RR}{\mathbb{R}}
\newcommand{\CC}{\mathbb{C}}

\newcommand{\diag}{\text{diag}}

\begin{document} 
\title{SHARP: a distributed, GPU-based ptychographic solver}

\ifx \aff \undefined

\author{Stefano Marchesini}
\affiliation{Lawrence Berkeley National Laboratory, Berkeley, CA, USA}

\author{Hari Krishnan}
\affiliation{Lawrence Berkeley National Laboratory, Berkeley, CA, USA}

\author{Benedikt J. Daurer} 
\affiliation{Uppsala University, Uppsala, Sweden}

\author{David A. Shapiro}
\affiliation{Lawrence Berkeley National Laboratory, Berkeley, CA, USA}

\author{Talita Perciano}
\affiliation{Lawrence Berkeley National Laboratory, Berkeley, CA, USA}

\author{James A. Sethian}
\affiliation{Lawrence Berkeley National Laboratory, Berkeley, CA, USA}

\author{Filipe R.N.C. Maia} 
\affiliation{Uppsala University, Uppsala, Sweden}

\else
\author[a]{Stefano Marchesini}{}
\author[a]{Hari Krishnan}{}
\author[b]{Benedikt J. Daurer}{}
\author[a]{David A. Shapiro}{}
\author[a]{Talita Perciano}{}
\author[a]{James A. Sethian}{}
\author[b]{Filipe R.N.C. Maia}{} 

\aff[a]{Lawrence Berkeley National Laboratory, Berkeley, CA, USA}
\aff[b]{Uppsala University, Uppsala, Sweden}
\fi

\date{\today}
\begin{abstract}  

  Ever brighter light sources, fast parallel detectors, and advances
  in phase retrieval methods, have made ptychography a practical and
  popular imaging technique.  Compared to previous techniques,
  ptychography provides superior robustness and resolution at the
  expense of more advanced and time consuming data analysis.  By
  taking advantage of massively parallel architectures,
  high-throughput processing can expedite this analysis and provide
  microscopists with immediate feedback.  These advances allow
  real-time imaging at wavelength limited
  resolution, coupled with a large field of view.  Here, we introduce
  a set of algorithmic and computational methodologies used at the
  Advanced Light Source, and DOE light sources packaged as a CUDA based
  software environment named \textsc{SHARP} (\url{http://camera.lbl.gov/sharp}), aimed at providing state-of-the-art high-throughput
  ptychography reconstructions for the coming era of diffraction
  limited light sources.
\end{abstract}
\preprint{LBNL-1003977}
\maketitle
\noindent

\section{Introduction} 

Reconstructing the 3D map of the scattering potential of a sample from
measurements of its far-field scattering patterns is an important
problem. It arises in a variety of fields, including optics
\cite{Fienup:1982,Luke_Burke_Lyon:2002}, astronomy \cite{Fienup:93},
X-ray crystallography \cite{Eckert:2012Xray100years}, tomography
\cite{Momose_Takeda_Itai_Hirano:1996}, holography
\cite{Collier_Burckhardt_Lin:1971,Marchesini_Boutet_Sakdinawat:2008}
and electron microscopy \cite{Hawkes_Spence:2007}. As such it has been
a subject of study for applied mathematicians for over a century.  The
fundamental problem consists of finding the correct phases that go
along with the measured intensities, such that together they can be
Fourier transformed into the real-space image of the sample. To help
recover the correct phases from intensity measurements a range of
experimental techniques have been proposed along the years, such as
interferometry/holography \cite{Collier_Burckhardt_Lin:1971}, random
phase masks \cite{Marchesini_Boutet_Sakdinawat:2008,Fannjiang_Liao:2012,Wang_Xu:2013}, gratings
\cite{Pfeiffer_Weitkamp_Bunk_David:2006}. A variety of numerical techniques
have also been recently developed, for example by approximating the problem as a matrix
completion problem \cite{Candes_Eldar_Strohmer_Voroninski:2013}, or by other convex relaxations  \cite{Waldspurger_dAspremont_Mallat:2013}
tractable by semidefinite programming.

Since its first demonstration \cite{miao:1999}, progress has been made in solving the phase problem for a single
diffraction pattern recorded from a non-periodic object, including the dynamic
update of the support \cite{marchesini:03} and a variety of projection algortihms
\cite{Bauschke_Combettes_Luke:2002,Marchesini:2007a,saddle}.
Such methods, referred to as coherent diffractive imaging (CDI), attempt to recover the complete
complex-valued scattering potential or electron density, and the complex exit wavefront scattered from the object, providing phase contrast
as well as a way to overcome depth-of-focus limitations of regular optical systems.

Ptychography, a relatively recent technique, provides the
unprecedented capability of imaging macroscopic specimens in 3D and
attain wavelength limited resolution along with chemical specificity
\cite{Rodenburg:2008}. Ptychography was proposed in 1969
\cite{Hoppe:1969,Hegerl_Hoppe:1970}, and later experimentally
demonstrated \cite{Nellist_McCallum_Rodenburg:1995,Chapman:1996}, with
the aim of improving the resolution in x-ray and electron
microscopy. Since then it has been used in a large array of
applications, and shown to be a remarkably robust technique for the
characterization of nano materials. A few software implementations of
the reconstruction algorithm exist such as ptypy
(http://ptycho.github.io/ptypy/) and PtychoLib \cite{Nashed2014}, and
a repository for sharing experimental data has been established
\cite{cxi}.

Ptychography can be used to obtain large high-resolution images. It
combines the large field of view of a scanning transmission microscope
with the resolution of scattering measurements. In a scanning
transmission microscope, operated in transmission mode, a focused beam
is rastered across a sample, and the total transmitted intensity is
recorded for each beam position. The pixel positions of the image
obtained correspond to the beam positions used during the scan, and
the value of the pixel to the intensity transmitted at that
position. This limits the resolution of the image to the size of the
impinging beam, which is typically limited by the quality of focusing
optics and work distance constraints. In ptychography, instead of only
using the total transmitted intensity, one typically records the
distribution of that intensity in the far-field, i.e. the scattering
pattern produced by the interaction of the illumination with the
sample. The diffracted signal contains information about features much
smaller than the size of the x-ray beam, making it possible to achieve
higher resolutions than with the scanning techniques. The downside of
having to use the intensities is that one now has to retrieve the
corresponding phases to be able to reconstruct an image of the sample,
which is made even more challenging by the presence of noise,
experimental uncertainties, and perturbations of the experimental
geometry. While it is a difficult problem, it is usually tractable by
making use of the redundancy inherent in obtaining diffraction
patterns from overlapping regions of the sample. This redundancy also
permits the technique to overcome the lack of several experimental
parameters and measurement uncertainties. For example, there are
methods to recover unknown illuminations
\cite{Thibault2008,Thibault2009,hesse2015,rank1}. As a testament to
their success these methods are even used as a way of characterizing
high quality x-ray optics \cite{Kewish2010,Honig:11,guizar-mirror},
the wavefront of x-ray lasers
\cite{ptychoxfel} and EUV lithography tools \cite{wojdyla2014ptychographic}.

Ptychographical phasing is a non-linear optimization problem 
\cite{fienup} still containing many open
questions \cite{synchroAP}. Several strategies, such as Alternating
Directions \cite{ZWen}, projections, gradient \cite{fienup}, conjugate
gradient, Newton \cite{chao,Thibault-Guizar,lbnlptycho}, spectral
methods\cite{marchesini2013augmented,synchroAP}
and Monte-carlo \cite{Maiden:2012}, have been proposed to handle
situations when both sample and positions
\cite{Maiden:2012,axel,fienup,marchesini2013augmented}, are unknown parameters in high dimensions, 
and to handle experimental situations such
as accounting for noise variance \cite{Thibault-Guizar,Godard:12}, partial coherence \cite{Fienup:93,Abbey2008,clark:2011,CDIpartialcoherence}\cite{clark:2011,marchesini2013augmented,tian2014multiplexed},
 background\cite{thurman2009,guizar:bias,chao,marchesini2013augmented} or vibrations.

Here, we describe an algorithm approach and software environment 
\textsc{SHARP}
(Scalable Hetereogeneous Adaptive Real-time Ptychography) that enables high 
throughput streaming analysis using computationally efficient phase retrieval
algorithms.
The high performance computational back-end written in C/CUDA and implemented for NVIDIA GPU architecures 
is hidden from the microscopist, but can be
accessed and adapted to particular needs by using a python interface
or by modifying the source code.

Using \textsc{SHARP} we have built an intuitive graphical user
interface that provides visual feedback, of both the recorded
diffraction data as well as the reconstructed images, throughout the
data aquisition and reconstruction processes at the Advanced Light Source (ALS).

We use a mathematical formulation of ptychography which was first introduced 
in  \cite{chao,lbnlptycho,ZWen,marchesini2013augmented,synchroAP}.

 \section{\textsc{SHARP} Software Environment}

 \subsection{Forward model}

\begin{figure} 
  \begin{center}
    \includegraphics[width=.75\columnwidth]{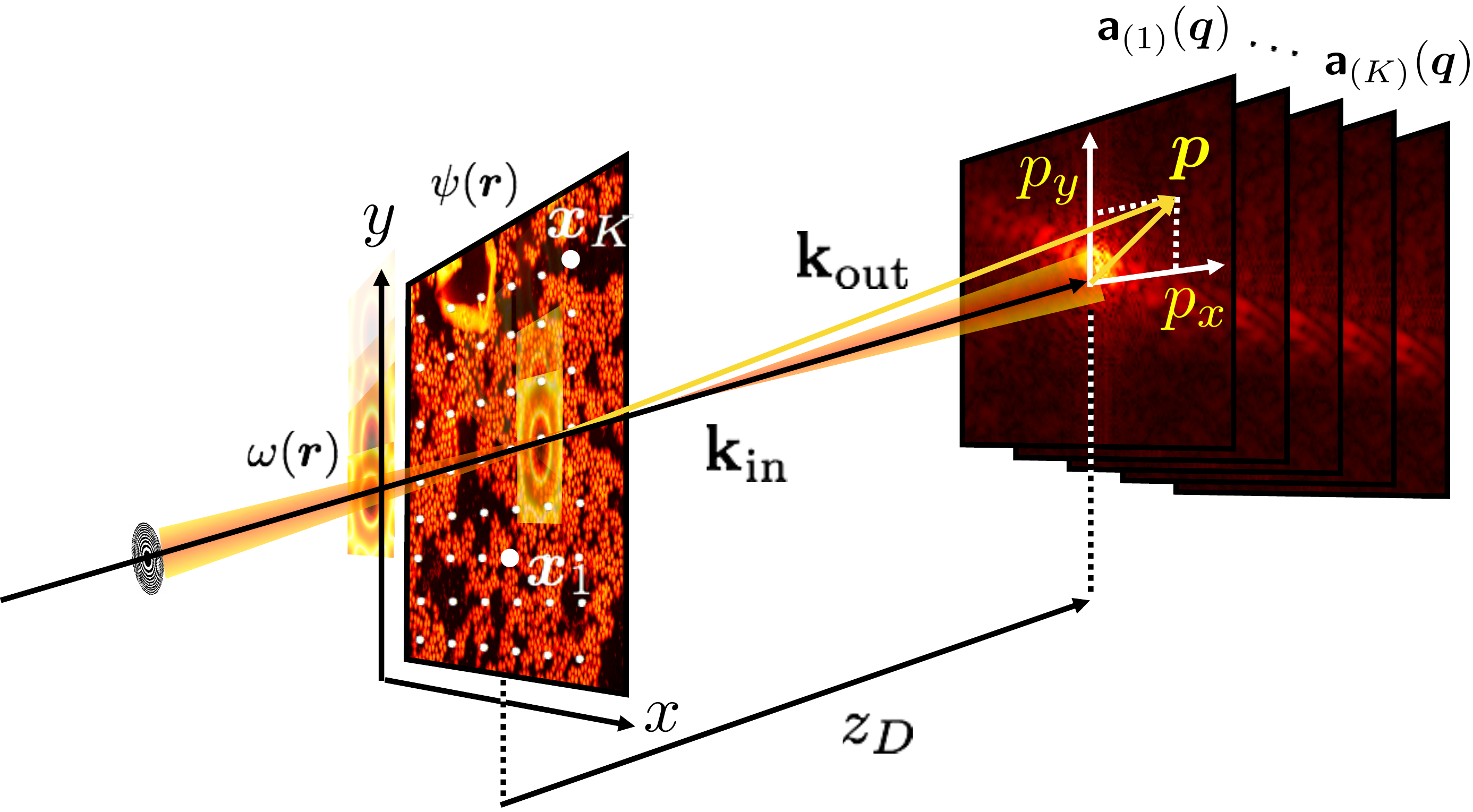}  
  \end{center}
  \caption{ 
  Experimental geometry in ptychography: an unknown sample with 
transmission $\psi(\vr)$ is rastered through an illuminating beam 
$\omega(\vr)$, and a sequence of diffraction measurements 
${\cal I}_{(i)}=|a_{(i)}(\vq)|^2$ are recorded on an area detector with pixel coordinates $\vp$ at a distance $z_D$ from the sample.}
\label{fig:expgeom}
\end{figure}

 In a ptychography experiment (see Fig. \ref{fig:expgeom}), one
 performs a series of diffraction measurement as a sample is rastered
 across an x-ray, electron or visible light beam. The illumination is formed by an x-ray optic
 such as a zone-plate. The measurement is performed by briefly
 exposing an area detector such as a CCD which records the scattered photons.

In a discrete setting, a two-dimensional small beam with distribution 
$w(\vr)$ of dimension $m_x \times m_y$ illuminates a subregion 
positioned at $\vx_{(i)}$ (referred to as frame) of an unknown 
object of interest $\psi(\vr)$ of dimension $n_x\times n_y$. Here $0<m< n$,
$i=1,\ldots,K$ and $K$ is the total number of frames (also referred to as ``views'' in the literature). 
For simplicity we consider square matrices. Generalization to non-square matrices is straightforward but requires more indices and
complicates notation. 

The pixel coordinates on a detector placed at a distance $z_D$ from the sample are described as $\vp = ( p_x, p_y,z_D)$.
Under far-field and paraxial approximations
the pixel coordinates are related to reciprocal space coordinates
\begin{align*}
  \vq = & \,\vk_{\mathrm{out}}-\vk_{\mathrm{in}}\\
            = & \,\frac{1}{\lambda} \, \left ({\tfrac{(p_x,p_y, z_D)}{\sqrt{p_x^2+p_y^2+z_D^2}} -(0,0,1) }\right) \\ 
            \simeq & \,  \frac{1}{\lambda  z_D} \, \left( p_x,p_y, 0 \right)
\end{align*}
where $\vk_{\mathrm{in}} = (0,0,k)$ and $\vk_{\mathrm{out}}= k \frac{\vp}{|\vp|}$ are the incident and scattered wave vectors that satisfy $|\vk_{\mathrm{in}} | = |\vk_{\mathrm{out}}| = k = 1/\lambda$, and $\lambda$ is the wavelength. With a distance $p_m$ from the center to the edge of the detector, the diffraction limited resolution (half-period) of the microscope is given by the lengthscale $r = \tfrac{\lambda z_D}{2 p_{m}}$. As a consequence, the coordinates in reciprocal and real space are defined as 
\begin{align*}
 &\vq = \left (\tfrac{\mu}{m r },\tfrac{\nu}{m r } \right),\,\,\mu,\nu\in\{0,\hdots, m-1\}
\end{align*}
and
\begin{align*}
 &\vr = \left (r\mu,r\nu \right),\,\, \mu,\nu\in\{0,\hdots, m-1\},\\
 &\vx_{(i)} = \left(r\mu',r\nu'\right),\,\,\mu',\nu'\in\{0,\hdots,n-m\}.
\end{align*}
While $\vx_{(i)}$ is typically rastered on a coarser grid, $\vr+\vx_{(i)}$ spans a finer grid of dimension $n\times n$.

In other words, we assume that a sequence of
$K$ diffraction intensity patterns ${\cal{I}}_{(i)}(\vq)$ are
collected as the
position of the object is rastered on the position $\vx_{(i)}$.
The simple transform ${a}_{(i)}=\sqrt{{\cal I}_{(i)} (\vq)}$ is a
variance stabilizing transform for Poisson noise
\cite{anscombe,makitalo}. The relationship among the amplitude
$a_{(i)}(\vq)$, the illumination function $w(\vr)$ and an unknown object $\psi(\vr)$
to be estimated can be expressed as follows:

\begin{align}
&a_{(i)}(\vq) = \left |{\cal F} w(\vr)\psi(\vr+\vx_{(i)}) \right |
\label{eq:data0}
\end{align}
and ${\cal F}$ is the two-dimensional discrete Fourier transform,
\begin{align}
&({\cal F} f )(\vq)=\tfrac{1}{\sqrt{m^2}} \sum_{\vr} e^{2\pi i\vq \cdot \vr} f(\vr).
\end{align}
where the sum over $\vr$ is given on all the indices $m \times m$ of $\vr$.
We define an operator $T_{(i)}$, that extracts a frame out of an image $\psi$, 
and build the illumination operator $\vQ_{(i)}$, which 
scales the extracted frame  point-wise by the illumination function $w$:
\begin{align*}
\vQ_{(i)}[ \psi](\vr)
  &= w(\vr)  \psi(\vr+\vx_{(i)}),\\
 &= w(\vr) T_{(i)}[ \psi](\vr),\\
&=z_{(i)}(\vr).
\end{align*}

With the operator $\vQ$, eq. (\ref{eq:data0}) can be represented compactly as:
\begin{eqnarray}
\label{eq:forwardproblem1}
&\va = |\vFQ\psi^\vee| \text{, or } 
\begin{cases}
\va = |\vF\vz|,\\
\vz = \vQ\psi^\vee,
\end{cases}
\end{eqnarray}
where the superscript $\psi^\vee$ denotes the linearized version of the image (the superscript will be omitted for simplicity), 
and more explicitely as:
\begin{eqnarray}
\label{eq:data}
&\overbracket{\left[\begin{array}{c} a_{(1)}\\ \vdots \\ a_{(K)}\end{array} \right]}^{\va\in \RR^{Km^2}}=\left |
\overbracket{
\left[\begin{array}{ccc} F & \ldots & 0 \\ \vdots & \ddots & \vdots\\ 0 & \ldots & F\end{array} \right]
}^{\vF\in \CC^{Km^2\times Km^2}}
\overbracket{
\left[\begin{array}{c} z_{(1)}\\ \vdots \\ z_{(K)}\end{array} \right] }^{\vz \in \CC^{K m^2}}
\right |, \\
\label{eq:overlap}
&\overbracket{\left[\begin{array}{c} z_{(1)}\\ \vdots \\ z_{(K)}\end{array} \right] }^{\vz \in \CC^{K m^2}}
=\overbracket{ \left[\begin{array}{c} \diag(w) T_{(1)}\\ \vdots \\ \diag(w)  T_{(K)}\end{array} \right]}^{\vQ \in \CC^{K m^2\times n^2},}
\overbracket{\left[\begin{array}{c} \psi_{1}\\ \vdots \\ \psi_{n^2}\end{array} \right] }^{\psi \in \CC^{n^2}}.
\end{eqnarray}
where  $\vz$ are $K$
frames extracted from the object $\psi$ and multiplied by the
illumination function $w$, and $\vF$ is the associated 2D DFT matrix
when we write everything in the stacked form
\cite{marchesini2013augmented}. 
When both the sample and the illumination are unknown,  
we can express the relationship (Eq. \ref{eq:overlap}) between the 
image $\psi$, the illumination $w$, and the frames $\vz$  in two forms:
\begin{eqnarray}
 \label{eq:standardprobe0} 
 \vz=\vQ \psi=\diag(\vS w) \vT \psi=\diag(\vT \psi)  \vS w
\end{eqnarray}
where $\vS\in \RR^{Km^2\times m^2}$ denotes the operator that replicates  the illumination $w$ into a stack of  $K$ frames,
since $\vQ \psi=\diag(\vS w) \vT \psi$ is the entry-wise product of $\vT\psi$ and $\vS w$. Eq. (\ref{eq:standardprobe0}) can be used to find $\psi$ or $w$ from $\vz$ and the other variable. 

The Fourier transform relationship used in equations (\ref{eq:data0}),
(\ref{eq:forwardproblem1}) and (\ref{eq:data}) is valid under
far-field and paraxial approximation, which is the focus of the current
release of \textsc{SHARP}. For experimental geometries such as Near Field\cite{stockmar2013near}, Fresnel \cite{vine2009ptychographic}, Fourier ptychography \cite{fourierptycho}, through-focus 
\cite{marrison2013ptychography} partially coherent multiplexed geometries 
\cite{tian2014multiplexed,batey2014information,dong2014spectral}, under-sampling conditions \cite{edo:pra13} and to account for noise variance \cite{marchesini2013augmented} , one can substitute the simple Fourier transform with the appropriate
propagator and variance stabilization \cite{chao}.

\subsection{Phase retrieval}
\label{sec:phase}
Projection operators form the basis of every iterative projection and projected gradient algorithms are implemented in \textsc{SHARP} and accessible through a library. The projection $P_{\va}$ ensures that the frames $\vz$ match the experiment, that is, they satisfy Eq. (\ref{eq:data}), and is referred to as data projector:
\begin{align}
\label{eq:Pa}
 P_{\va} \vz=\vF^{*}\frac{\vF\vz}{|\vF\vz|}\va
\end{align}
while the projection $P_{\vQ}$ onto the range of $\vQ$ (see Fig. \ref{fig:parallel}):
\begin{align}
\label{eq:Pq}
P_{\vQ}=\vQ(\vQ^*\vQ)^{-1}\vQ^*
\end{align}
ensures that overlapping frames $\vz$ are consistent with each other and satisfy Eq. (\ref{eq:overlap}).

The projector $P_{\va}$ is relatively robust to Poisson noise
 \cite{anscombe}, but weighting factors to account for noisy pixels can be easily added \cite{lbnlptycho}.

 Using relationship (\ref{eq:standardprobe0}), we can update the image $\psi$ from $w$ and frames $\vz$:
\begin{eqnarray}
\label{eq:standardobject}
\psi\leftarrow &
\frac{ \vQ^\ast \vz}
{\vQ^\ast \vQ}
\end{eqnarray}
or the illumination $w$ from an image $\psi$ and frames $\vz$ \cite{Thibault2008,Thibault2009}
multiplying (Eq. \ref{eq:standardprobe0}) on the left by $\diag{\vT \bar \psi}$ and solving for $w$:
\begin{eqnarray} 
\label{eq:standardprobe}
w\leftarrow&{\frac{\vS^\ast \diag( \vT \bar \psi ) \vz}{\vS^\ast \vT |\psi|^2 }},
\end{eqnarray}S
where $\bar \psi$ denotes the complex conjugate of $\psi$. See
\cite{rank1} for alternative updates, and \cite{hesse2015} for
convergence theory behind a similar blockwise optimization strategy.
Several possible pathologies need to be accounted for when updating
both $\psi$ and $w$:
\begin{itemize}
\item{Combined drift of the illumination and  the image in real space. 
 Drift is eliminated by keeping the illumination in the center of the frame by computing the center of mass and correcting for drifts after every  update of the illumination.}
\item{Fourier space drifts and grid pathologies are supressed by enforcing
  either the absolute value $a_w=|{\cal F} w_0|$ or support $m_w$ of the Fourier transform of the unknown illumination $w_0$.}
\item{A possible global phase factor between the solution and the reconstruction
  is taken into account in the error calculation.}
\end{itemize}

A typical reconstruction with \textsc{SHARP}  uses the following sequence:
\begin{framed}
\fontsize{8}{9.2}\selectfont

\begin{enumerate}
\item Input data ${\cal I}(\vq)$, 
 translations $\vx$. Optional inputs: initial image $\psi^{(0)}$, illumination
 $w^{(0)}$, illumination Fourier mask $m_{w}$ and illumination Fourier amplitudes
 $a_{w}$.
\item If $w^{(0)}$ is not provided, initialize illumination by setting $w^{(0)}$
  to the inverse Fourier transform of the square root of the average frame.
\item If $\psi^{(0)}$ is not provided, initialize the image by filling
  $\psi^{(0)}$ with random numbers uniformly drawn from $[0,1)$.
\item Build up  $\vQ$, $\vQ^\ast$, and $(\vQ^\ast \vQ)^{-1}$, and frames $\vz^{(0)}=\vQ \psi^{(0)}$;
\item \label{item:iterate} Update the frames $\vz$  according to \cite{raar}
using projector operators defined in (Eqs. (\ref{eq:Pa},\ref{eq:Pq}))
below:
$$
\vz^{(l)}:= [2\beta P_{\vQ}P_{\va}+(1-2\beta)  P_{\va} +\beta(P_{\vQ}-I)]\vz^{(l-1)},
$$
where $\beta\in(0.5,\, 1]$ is a scalar factor set by the user (set to 0.75 by
  default, which works in most cases).
\item Update image $\psi^{(\ell)}$ using using Eq. \ref{eq:standardobject}.
\item If desired, compute a new illumination $w$ using Eq. \ref{eq:standardprobe}. If
  $m_{w}$ is given apply the illumination Fourier mask constraint:
$$
w^{(\ell)} := {\cal F}^{-1}\{({\cal F}w) m_{w}\},
$$
else if $w_{I}$ is given apply the illumination Fourier intensities constraint:
$$
w^{(\ell)} := {\cal F}^{-1}\left\{\frac{{\cal F}w}{|{\cal F}w|} a_{w}\right\},
$$
else simply keep the unconstrained illumination $w^{(\ell)} := w$.

Now compute center of mass of $w^{(\ell)}$ and shift it to fix the translation of the object.
\item If desired do background retrieval, that is, estimate static background and remove it in the iteration as described in \cite{marchesini2013augmented} (p.7, Eq. 30).
\item Iterate from \ref{item:iterate} until one of the metrics from
  Eqs. \ref{eq:amplitudemetric},\ref{eq:overlapmetric},\ref{eq:solutionmetric} drops below a user defined level or untill a maximum iteration for time-critical applications, and return $\psi^{(\ell)}$ and $w$.
\end{enumerate}
\end{framed}

The  metrics $\varepsilon_\Delta$, $\varepsilon_a, \varepsilon_Q, \varepsilon_\Delta$ used to monitor
progress amd stagnation are the {\it normalized mean square root error} (nmse) from the corresponding projections of $\vz$:
\begin{eqnarray}
\label{eq:amplitudemetric}
\varepsilon_{\va} \left(\vz\right)&=&\tfrac {\left \| \left [ P_{\va} - I\right ] \vz \right \|}{ \| \va\| }
,\\
\label{eq:overlapmetric}
 \varepsilon_Q\left (\vz\right)&=&\tfrac {\left \| \left [ P_Q -I \right ]\vz  \right \|}{ \| \va\|},
\\
\label{eq:stagnationmetric}
\varepsilon_{\Delta } \left(\vz^{(l)},\vz^{(l-1)}\right)&=&\tfrac {\left \| \vz^{(l)}-\vz^{(l-1)} \right \|}{ \| \va\| }
\end{eqnarray}
where $I$ is the identity operator, and $\vz^{(0)}=0$,

 For benchmarking purposes,  when using a simulation from a known solution $\psi_0$, 
the following metric can also be used:
\begin{eqnarray}
\label{eq:solutionmetric}
  \varepsilon_0\left (\vz \right )&=& \tfrac 1 
{\|\vQ^\ast \vz_0 \|}   
{ \min_\varphi \left  \| \vQ^\ast ( e^{i\varphi} \vz- \vz_0) \right \|},
\end{eqnarray}
where $\varphi$ is an arbitrary global phase factor, and 
$\vz_0=\vQ \psi_0$. Notice the additional scaling factor $\vQ^\ast$ used in $\varepsilon_0$.


The initial values for the input data and translations can either be
loaded from file or set by a python interface. The starting
``zero-th'' initial image is loaded from file, set to a random image,
or taken as a constant image.

\subsection{Computational Methodology}

\begin{figure}
  \begin{center}
    \includegraphics[width=.65\columnwidth]{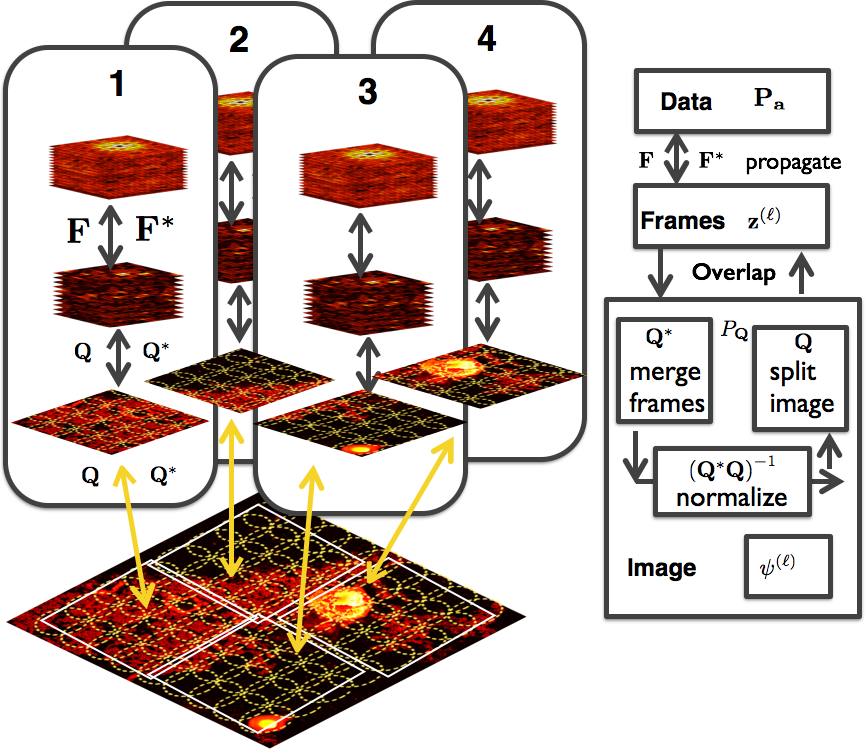}  
  \end{center}
  \caption{Schematic of the ptychographic reconstruction algorithm implemented in \textsc{SHARP}. 
    The iterative reconstruction scheme is shown on the right.
    To achieve the highest possible throughput and scalability
    one has to parallelize across multiple GPUs as shown on the left for the case of 4 GPUs. 
    As most ptychographic
    scans use a constant density of scan point across the object, we
    expect to be able to achieve a very even division, resulting in
    good load balancing.  \textsc{SHARP} enforces an overlap
    constraint between the images produced by each of the GPUs, and
    also enforces that the illumination recovered on each GPU agree
    with each other. This is done by default at every iteration.    
    } 
   \label{fig:parallel}
\end{figure}

\textsc{SHARP} was developed to achieve the highest performance,
taking advantage of the algorithm described earlier and using
a distributed computational backend.  The ptychographic reconstruction
algorithm requires one to compute the product of several linear
operators ($\vQ$, $\vQ^\ast$,  $\vF$,  $\vF^\ast$,
$\vS$, $\vS^\ast$) on a set of frames $\vz$, an image $\psi$ and an
illumination $w$ several times. We use a distributed GPU architecture
across multiple nodes for this task (Fig. \ref{fig:parallel}).

To implement fast operators, a set of GPU kernels and MPI
communication are necessary. The split ($\vQ\psi$) and overlap ($\vQ^\ast\vz$)
kernels are among the most bandwidth demanding kernels and play an important role in the process.

The strategy used to implement those kernels impacts directly the
overall performance of the reconstruction algorithm.  To divide the
problem among multiple nodes, \textsc{SHARP} initially determines the size of
the final image based on the list of translations, frames size, and
resolution. It subsequently assigns a list of translations to every
node and loads the corresponding frames onto GPUs. 

The split ($\vQ
\psi$) and FFT ($\vF$) operations are easily parallelized because of
the framewise intrinsic independence.  Summing the frames onto an
image ($\vQ^\ast \vz$) requires a reduction for every image pixel
across neighboring MPI nodes. Within each GPU the image is divided into blocks
and we first determine which frames contribute to each block. The
contributing frames are summed and then the resulting image is summed
across all MPI nodes. We use shared memory or constant memory,
depending on GPU compute capability, to store frame translations, and
we use kernel fusion to reduce access to global memory. The last step
of summing across all MPI nodes does not necessarily have to be done
at every iteration, at the cost of slower convergence
\cite{liu2015asynchronous}, but that is the
default.

Timing to compute the overlap at each iteration depends on the size of
the image and number of frames on top of each pixel, i.e. the density
but not the size of the frames.
 
In addition to the high performance ptychographic algorithm, the
\textsc{SHARP} software environment provides a flexible and modular
framework which can be changed and adapted to different needs.
Furthermore, the user has control of several options for the
reconstruction algorithm, which can be used to guarantee a balance
between performance and quality of the results. These include the choice of
illumination Fourier mask, illumination Fourier intensities and the $\beta$
parameter, as well as how often to do different operations such as
illumination retrieval, background retrieval, and synchronization of the
different GPUs. For more details  we refer the reader to the documentation 
(http://www.camera.lbl.gov/sharp).

\section{Applications and Performance}
\begin{figure}
  \begin{center}
    \includegraphics[width=.75\columnwidth]{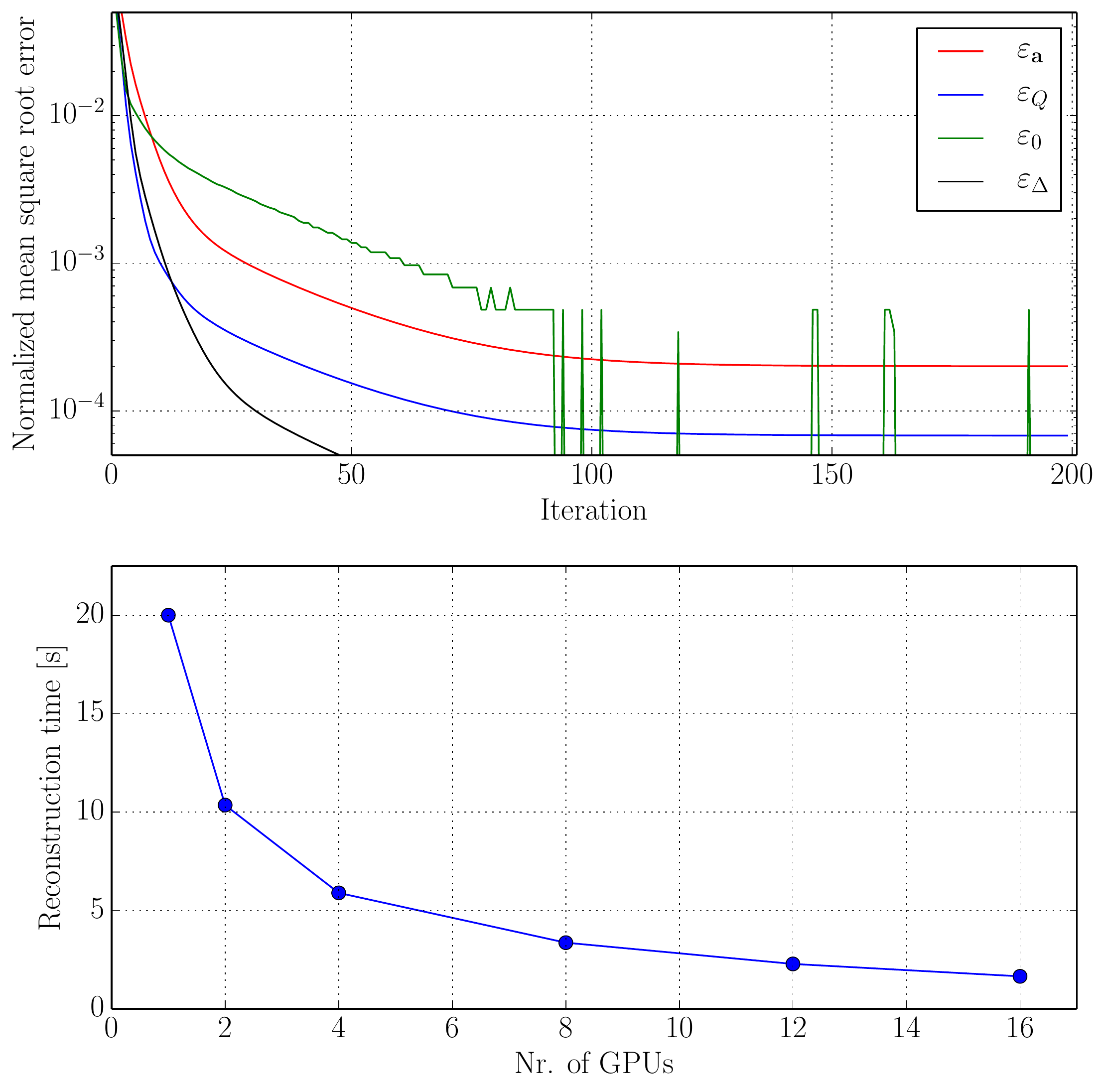} 
 \end{center}
 \caption{
Convergence rate (top) per iteration and timing (bottom) to process 10,000 frames of dimension $128 \times 128$ extracted from an 
image of size $1000\times 1000$ as a function of the number of nodes. All residuals decrease rapidly; numerical precision limits the (weighted) comparison with the known solution $\varepsilon^\prime(\vz)$.  Reconstruction  is achieved (${\varepsilon_0 < 5e-4}$) in under 2  seconds  using a cluster with 4 compute nodes with 4 GTX Titan GPU per node (16 total, 43000 cores), 96 GBytes GPU memory, 1 TByte RAM, and 24 TBytes storage, infiniband.
Timing contributions for corresponding computational kernels are 
$(\vQ^\ast \vQ)^{-1} \vQ^\ast$ 30 \%, $\vF,\vF^\ast$ 20 \%, $\vQ$ 20 \%, $\vS^=ast$ 5 \%,
elementwise operations 20 \%, and residual calculation 5 \%. No illumination
retrieval was done, as the exact illumination was given. The simulation was done using periodic
boundary conditions to avoid edge effects. 
}
\label{fig:timing}
\end{figure}

\textsc{SHARP} enables high-throughput streaming analysis using
computationally efficient phase retrieval algorithms.  In this section
we describe a typical dataset and sample that can be collected in less
than 1 minute at the ALS, and the computational backend to provide
fast feedback to the microscopist.  

To characterize our performance, we use both simulations and experimental data.  We use simulations to
compare the convergence of the reconstruction algorithm to the ``true
solution'' and characterize the effect of different light sources,
contrast, scale, noise, detectors or samples for which no data exists
yet. 

Experimental data from ALS used to characterize battery
materials, green cement, magnetic materials, at different wavelengths and orientation has
been successfully reconstructed
\cite{shapiro2014chemical,yu2015dependence,bae2015soft,li2015electrode,shi2016soft}
using the software described in this article.  

We also describe a
streaming example in which a front-end that operates very close to the
actual experiment sends the data to the reconstruction backend that
runs remotely on a GPU/CPU cluster. Further details about the
streaming front-end and processing back-end pipeline will be published in 
an upcoming paper by our group.

\subsection{Simulations and performance}
As a demonstration, we start  from a sample that was composed of
colloidal gold nanoparticles of 50 nm and 10 nm deposited on a
transparent silicon nitride membrane. An experimental image was obtained by scanning
electron microscopy, which provides high resolution and contrast but
can only view the surface of the sample.  

We simulate a complex transmission function by scaling the image amplitude from 0 to 50 nm
thickness, and using the complex index of refraction of gold at 750 eV
energy from [henke.lbl.gov].  The illumination is generated by
a zone-plate with a diameter of 220 microns and 60 nm outer zone width,
discretized into (128$\times$128) pixels in the far field.

\subsection{Experimental example }
 
\begin{figure}
  \begin{center}
  	\includegraphics[width=.85\columnwidth]{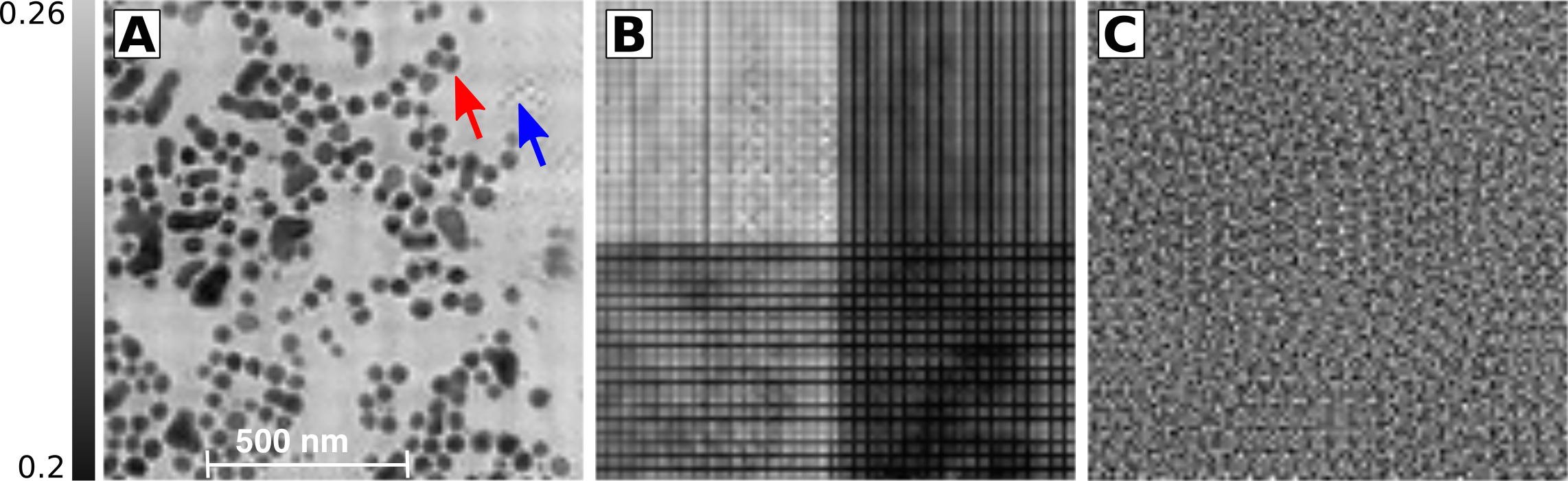} 
 \end{center}
 \caption{
Reconstruction of a test sample consisting of gold balls with
diameters of 50 and 10 nm. Detector pixel size 30 microns, $1920\times960$ pixels $80$ mm downstream from the sample, cropped and downsampled to $128$, scan of (50$\times$ 50) points,  illumination is generated by
a zone-plate with a diameter of 220 microns and 60 nm outer zone width.  A) Phase image generated by \textsc{SHARP} using
the algorithm described in section \ref{sec:phase} applying the illumination Fourier mask constraint 
and turning on background retrieval. The red arrow points to a
collection of 50 nm balls while the blue arrow points to a collection
of 10 nm balls.  The pixel size is 10 nm.  B) Same as (A) except without
enforcing the illumination Fourier mask. C) Same as (A) but without using the
background retrieval algorithm.  }
\label {fig:gold_balls}
\end{figure}

Figure \ref{fig:gold_balls} shows ptychographic reconstructions of a
dataset generated from a sample consisting of gold balls with
diameters of 50 and 10 nm.  The data were generated using 750 eV
x-rays at beamline 5.3.2.1 of the Advanced Light Source, with high
stability position control of the soft x-ray scanning transmission
microscope. Exposure time was 1 second and the dataset consists of a
square scan grid with 40 nm spacing ( see \cite{shapiro2014chemical} for details of the experimental setup).  The reconstructions consisted of 300 iterations
of the RAAR algorithm with a illumination retrieval and background retrieval
step every other iteration. The initial illumination is generated by 
 (1) computing the average of the measurements, (2) seting everything below a threshold to 0, and everything above a threshold to a constant average value (3) applying an inverse FFT.  The image is initialized with complex independent identically distributed (i.i.d.) pixels, or a constant average value.
 

\subsection{Interface and Streaming }

Common processing pipelines used for ptychographic experiments usually
have a series of I/O operations and many different components
involved. We have developed a streaming pipeline, to be deployed at the COSMIC
beamline at the ALS, which allows users to monitor
and quickly act upon changes along the experimental and computational pipeline.

\begin{figure}
  \begin{center}
    \includegraphics[width=.85\columnwidth]{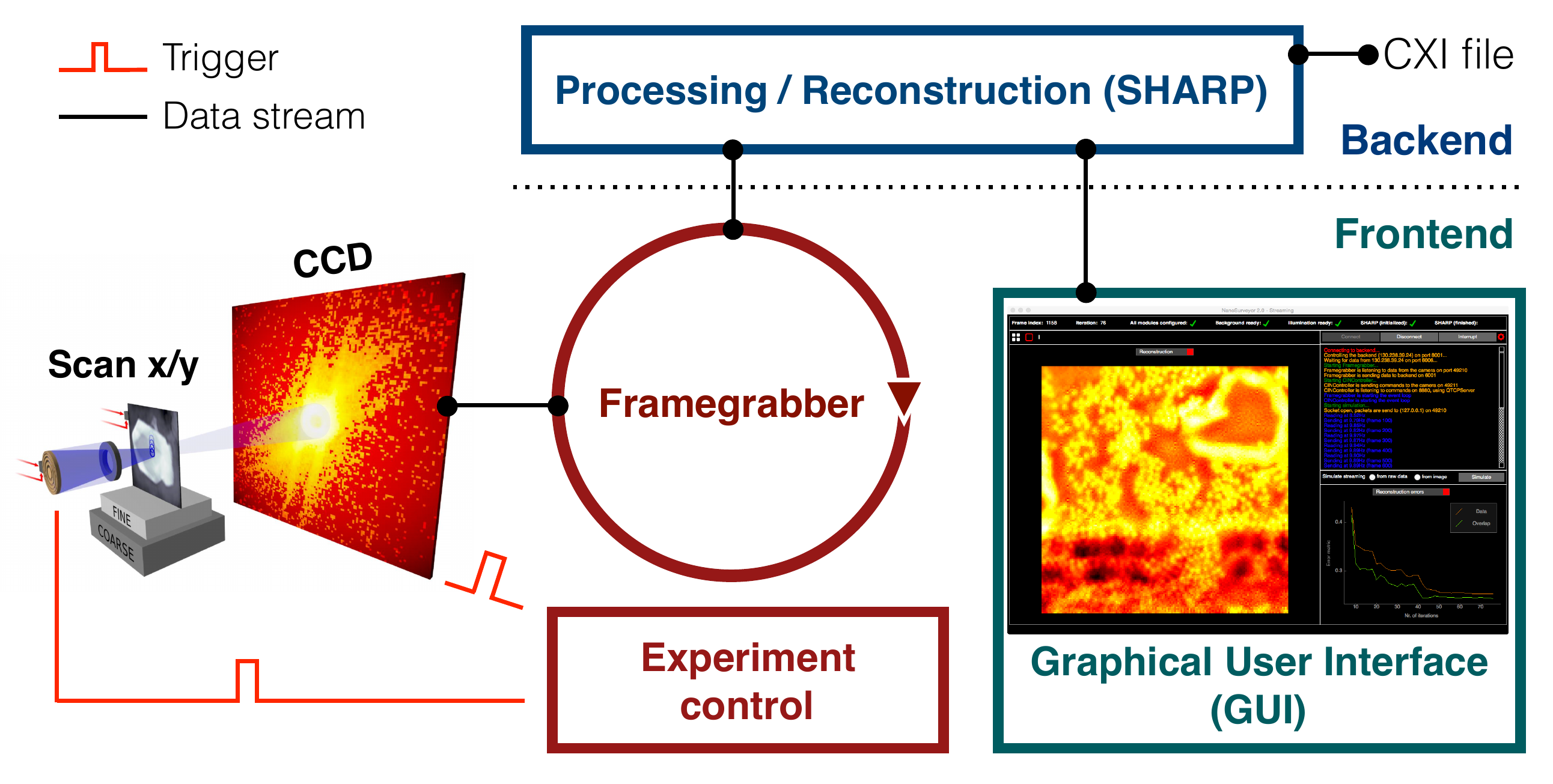} 
  \end{center}
 \caption{
Overview of the
components involved in the software structure of the streaming
pipeline. In order to maximize the performance of this streaming
framework, the frontend operates very close to the actual experiment,
while the backend runs remotely on a powerful GPU/CPU cluster.
 As soon as diffraction data is recorded by the
CCD camera, a live view of the ptychographic reconstruction is transmitted to the Graphical User Interface,
and  the user is able to control  and monitor (top panel) the 
current status of the data streams and analysis,  (bottom right panel) .
}
\label{fig:streaming_pipeline}
\end{figure}

The streaming pipeline is composed of a front-end and a back-end
(Fig.~\ref{fig:streaming_pipeline}).  The front-end consists of a Graphical User Interface (see Fig. \ref{fig:streaming_pipeline}),
 a worker that grabs frames from
the detector, and an interface that monitors network activity, experimental
parameters (position, wavelength, exposure, etc...), and provides a live view 
of the ongoing reconstruction.


On the back-end side, the streaming infrastructure is composed of a communication
handler and a collection of workers addressing different tasks such as dark calibration, 
detector correction, data reduction, ptychographic reconstruction and writing output to file.

This software architecture allows users an intuitive, flexible and responsive
monitoring and control of their experiments. Such a tight integration between
data aquisition and analysis is required to give users the feedback they expect
from a STXM instrument.

\section{Conclusions}

In this paper we described \textsc{SHARP}, a high-performance software environment for 
ptychography reconstructions, and its application as part of quick
feedback system used by the ptychographic mircoscopes installed at the Advanced 
Light Source. 

Our software provides
a modular interface to the high performance computational
back-end and can be adapted to different needs. 
Its fast throughput provides near real time feedback to microscopists and this
also makes it suitable as a corner stone for demanding higher dimensional analysis such as spectro-ptychography or tomo-ptychography.

With the coming new generation light sources and faster detectors, the
ability to quickly analyse vast amounts of data to obtain large
high-dimensional images will be an enabling tool for science. 

\section{Acknowledgments}

We acknowledge useful discussions with Chao Yang, H-T Wu, J. Qian and
Z. Wen.  This work was partially funded by the Center for Applied
Mathematics for Energy Research Applications, a joint ASCR-BES funded
project within the Office of Science, US Department of Energy, under
contract number DOE-DE-AC03-76SF00098, by the Swedish Research Council
and by the Swedish Foundation for Strategic Research.

\ifx \aff \undefined
\else
 \bibliographystyle{iucr}    
\fi

\bibliography{ptycho_probe}

\end{document}